\documentstyle[aps,prb]{revtex}

\def\SSC{Solid State Commun.\ }
\def\prep{Phys. Rep.\ }
\def\ZETF{Zh. Eksp. Teor. Fiz.\ }
\def\SPJ{Sov. Phys.-JETP\ }
\def\WRM{Waves Random Media\ }

\begin{document}

\draft

\title{Conductance of Two-Dimensional Imperfect Conductors:\\
Does the Elastic Scattering Preclude from Localization at $T=0$?}

\author{Yu.~V.~Tarasov}

\address{Institute for Radiophysics \& Electronics,
National Academy of Sciences of Ukraine,\\
12 Acad. Proskura Str., Kharkov 310085, Ukraine}

\date{\today}

\maketitle

%%%----------------------------------------------------------------
\begin{abstract}
The elastic electron-impurity scattering is proven analytically to prevent
from interferential localization in 2D wires with more then one conducting
channel. Unconventional diffusive regime is found in the length region
where the electrons are usually considered as localized. Ohmic dependence 
of $T=0$ conductance is predicted instead of exponential, with 
length-dependent diffusion coefficient.

\end{abstract}
%%%-----------------------------------------------------------------


%%%%%%%%%%%%%%%%%%%%%%%%%%%%%%%%%%%%%%%%%%%%%%%%%%%%%%%
% \narrowtext

Electronic and classical-wave transport in random systems of various
dimensionalities have been attracting much attention for scores of years.
Numerous attempts in this field are concentrated around the problem of
Anderson localization whose aspects build up to a great extent the
understanding of metal-insulator transitions. Prospects of researches in
this area are substantially determined by the claims of the one-parameter
scaling theory of localization.\cite{abrah79}  Although universality of the
one-parameter scaling was quite long
challenged,\cite{abrikos81,kravts_lerner84} attempts were made (and still
persist) to improve the scaling approach for its relative convenience and
simplicity.\cite{shreib_ottom92,dobrosavl97} They were stimulated
considerably by experimental findings of unexpectedly anomalous transport
in dilute two-dimensional (2D) electron and hole systems.\cite{kravch9596}
Also unconventional experimental results have stimulated the development of
different approaches to the problem of quantum transport in disordered 2D
systems (see, e.g., discussion in  Ref.~\onlinecite{aag99}), among which
the most intriguing expectations are associated with the Coulomb
interaction of carriers.\cite{shepelyan94,weinman97} Yet the transport
theories with {\em e-e} interaction still cannot claim for the general
acceptance because of substantial controversy in interpreting the role of
the interaction within different domains of parameters corresponding to
diffusive\cite{berkovits96} and localized\cite{efros95,talamantes96}
regimes.

In spite of an ample variety of theoretical approaches to the problem of
localization, some points in this field are still vague, and therefore
attract intensive researches. The important obstacle is insufficient
mathematical grounds for localization in 2D and 3D random systems, as
opposed to rigorous results\cite{LifGredPas} for 1D systems with
arbitrary strength of disorder. Meanwhile, it is instructive to point out
that elaboration of practical asymptotic methods for
calculating the disorder-averaged many-particle characteristics
(conductivity, density-density correlator, etc), leaving out the profound
spectral analysis, was even more important for the theory of 1D random
systems than development of the mathematical
foundation.\cite{berezinski73,abrikryzh78,kanercheb87,kanertar88} Arguments
of a comparable standard in favour of localization, as well as against it,
for 2D and 3D systems have not been found yet, except for some aspects of
weak localization problem.\cite{altaronkhmel82,altarkhmellark82}

The objective of this paper is to derive a quantum theory, i.e. based
on waveguide ideology, well adapted for the analysis of charge transfer in
weakly disordered 2D conductors. This theory can be built with the same
rigor as that in
Refs.~\onlinecite{berezinski73,abrikryzh78,kanercheb87,kanertar88} due to
the problem of quantum transport in 2D {\em waveguide systems} being
reduced exactly to a set of purely one-dimensional subsidiary problems.
Although substantial complication of the potentials arises as a requital
for such a reduction (the problems turn out to be non-Hermitian), the
dynamic properties of one-dimensional systems are now subject to the
canonical analysis beyond the scaling hypothesis, RMT, etc.

We consider a two-dimensional rectangular sample of the length $L$ in $x$ 
direction and the width $D$ in $y$ direction, where non-interacting 
electrons subject to static random potential are confined between the 
hard-wall lateral boundaries $y=\pm D/2$ while in the direction of current 
($x$) we suppose the system open. The dimensionless conductance $g(L)$ (in 
units $e^2/\pi\hbar$) is computed from the linear response 
theory,\cite{kubo57} whence at zero temperature the formula follows

\begin{equation}
g(L)=
- \frac{4}{L^2} \int\!\!\!\!\int_L dx    % \limits_{\!\!(L)}\!dx
dx'\sum_{n,n'=1}^{\infty}
\frac{\partial G_{nn'}(x,x')}{\partial x}
\frac{\partial G_{nn'}^*(x,x')}{\partial x'} \ .
\label{Cond-mode}
\end{equation}
Here $G_{nn'}(x,x')$ is the retarded one-electron Green function in the
coordinate-mode representation, i.e. Fourie-transformed over the transverse
coordinate. This function obeys the equation

\begin{equation}
\bigg[ \frac{\partial^2}{\partial x^2}+k_n^2+i0
- V_n(x)\bigg]G_{nn'}(x,x')
-\sum_{m=1\atop(m\neq n)}^{\infty} U_{nm}(x)G_{mn'}(x,x')
=\delta_{nn'}\delta(x-x') \ ,
\label{ModeEq}
\end{equation}
where $k_n^2=k_F^2-(n\pi/D)^2$ is the longitudinal mode energy, $k_F$ is the
Fermi wavenumber, $U_{nm}(x)$ is the mode matrix element of the ``bulk''
random potential $V(\bbox{r})$ which is assumed to have zero mean and the
binary correlator

\begin{equation}
\left< V(\bbox{r})V(\bbox{r}') \right>={\cal Q}W(\bbox{r}-\bbox{r}') \ ,
\label{VrVr}
\end{equation}
$\bbox{r}=(x,y)$. The angular brackets in Eq.~(\ref{VrVr}) stand for
impurity averaging, the function $W(\bbox{r})$ is normalized to unity and
has the correlation radius $r_c$.

From the technical point of view it is important that the diagonal matrix
element $V_n(x)\equiv U_{nn}(x)$ is initially separated in
Eq.~(\ref{ModeEq}) from off-diagonal elements, so that the matrix
$\|U_{nm}\|$ governs inter-mode transitions only. This enables to reduce
strictly the problem of finding overall set of the functions
$G_{nn'}(x,x')$ to the solution of a subset of purely one-dimensional
closed equations for the diagonal mode functions $G_{nn}(x,x')$. The exact
``one-dimensionalization'' procedure is sketched out below.

First, we introduce the auxiliary (trial) Green function $G_n^{(V)}(x,x')$
obeying the equation

\begin{equation}
\left[\frac{\partial^2}{\partial x^2}+k_n^2+i0- V_n(x)\right]
G_n^{(V)}(x,x')=\delta(x-x')
\label{GVn}
\end{equation}
and Sommerfeld's radiative conditions\cite{BassFuks79} at the strip
ends $x=\pm L/2$, which seem natural for an open system. Then, turning from
Eq.~(\ref{ModeEq}) to the consequent integral equation

\begin{equation}
G_{nn'}(x,x')=G_n^{(V)}(x,x')\delta_{nn'}
+\sum_{m=1}^{\infty}\int_L dt\,
{\sf R}_{nm}(x,t)G_{mn'}(t,x')
\label{LippShwin}
\end{equation}
with the kernel
\begin{equation}
{\sf R}_{nm}(x,t)=G_n^{(V)}(x,t)U_{nm}(t) \ ,
\label{sfR}
\end{equation}
one can express all the off-diagonal mode elements $G_{mn}$ via the
diagonal ones $G_{nn}$ by means of the linear operator $\hat{\sf K}$,

\begin{equation}
G_{mn}(x,x')=\int_Ldt\,{\sf K}_{mn}(x,t)G_{nn}(t,x') \ .
\label{G_mn-sol}
\end{equation}
The equation for the matrix elements ${\sf K}_{mn}(x,x')$ of $\hat{\sf K}$
results directly from Eq.~(\ref{ModeEq}),

\begin{equation}
{\sf K}_{mn}(x,x')={\sf R}_{mn}(x,x')
+\!\!\sum_{k=1\atop (k\neq n)}^{\infty}
\int_Ldt\,{\sf R}_{mk}(x,t){\sf K}_{kn}(t,x') \ .
\label{K_mn}
\end{equation}
This equation belongs to a class of multi-channel Lippmann-Schwinger 
equations that are known to be extremely singular in general, in contrast 
to their single-channel counterparts.\cite{Taylor72} However by choosing 
the trial Green function $G_n^{(V)}$ as a zero approximation for $G_{nn'}$ 
and perturbing it by the inter-mode potentials $U_{nm}(x)$ only, we manage 
to avoid the above mentioned singularity.  Therefore the solution of 
Eq.~(\ref{K_mn}) can be written in the form

\begin{equation}
\hat{\sf K}=\left(\openone-\hat{\sf R}\right)^{-1}\hat{\sf R}{\bf P}_n \ ,
\label{hatK}
\end{equation}
where $\hat{\sf R}$ is an operator acting in the mixed coordinate-mode space
($x,n$) and specified by the matrix elements (\ref{sfR}). It is important
that the indicated space contains all the waveguide modes except for the
$n$-th mode itself. The projection operator ${\bf P}_n$ makes the mode
index of any operator that stands next to ${\bf P}_n$ (both from the left
and right) equal to $n$. 

From Eqs.~(\ref{ModeEq}), (\ref{G_mn-sol}),
(\ref{hatK}) we obtain the exact closed one-dimensional equation for each
diagonal function $G_{nn}(x,x')$ separately,

\begin{equation}
\left[\frac{\partial^2}{\partial x^2}+\kappa_n^2
+i0-V_n(x)-\Delta{\hat{\cal T}}_n\right]
G_{nn}(x,x')=\delta(x-x') \ ,
\label{1Deq}
\end{equation}
with $\kappa_n^2=k_n^2-\langle{\hat{\cal T}}_n\rangle$,
$\Delta{\hat{\cal T}}_n={\hat{\cal T}}_n-\langle{\hat{\cal T}}_n\rangle$.
The operator $\Delta{\hat{\cal T}}_n$ acts on the variable $x$ only since
from Eq.~(\ref{hatK}) it follows that the operator ${\hat{\cal T}}_n$ is
a two-dimensional $T$-matrix\cite{Taylor72} enveloped by the projective
operators ${\bf P}_n$,

\begin{equation}
{\hat{\cal T}}_n={\bf P}_n\hat{\cal U}\left(\openone-
\hat{\sf R}\right)^{-1}\hat{\sf R}{\bf P}_n =
{\bf P}_n \hat{\cal U}\left(\openone-
\hat{\sf R}\right)^{-1}{\bf P}_n \ ,
\label{Tn}
\end{equation}
$\hat{\cal U}$ is the intermode scattering operator in ($x,n$) space,
specified by matrix elements $\left<x,k\right|\hat{\cal
U}\left|x',m\right>=U_{km}(x)\delta(x-x')$.
Hereinafter, when analyzing Eq.~(\ref{1Deq}), we regard a set of the
renormalized energies $\kappa_n^2$ ($n=1,2,\ldots$) as representing the new
``unperturbed spectrum'' of the system, instead of the primordial spectrum
$\{k_n^2\}$.  The perturbation theory will then be developed making use of
the appropriate zero-mean potentials $V_n(x)$ and $\Delta{\hat{\cal T}}_n$.

To complete the one-dimensionalization we express the conductance
(\ref{Cond-mode}) through the diagonal Green functions $G_{nn}$ and the
trial functions $G^{(V)}_n$ (both one-dimensional!). In this paper we
focus on the case of {\em weak electron-impurity scattering} specified by
the inequalities

\begin{equation}
k_F^{-1},r_c\ll\ell \ ,
\label{weakscat}
\end{equation}
with $\ell=2k_F/{\cal Q}$ denoting a {\it semiclassical} mean free path
evaluated for $\delta$-correlated 2D random potential, i.e.
$W(\bbox{r})=\delta(\bbox{r})$ in Eq.~(\ref{VrVr}). The conditions
(\ref{weakscat}) allow to expand the operator $\hat{\sf K}$,
Eq.~(\ref{hatK}), to lowest order in the inter-mode operator $\hat{\sf R}$,
what in turn enables to replace the exact operator ${\hat{\cal T}}_n$
from Eq.~(\ref{Tn}) by its approximate value

\begin{equation}
{\hat{\cal T}}_n\approx {\bf P}_n\hat{\cal U}{\hat G}^{(V)}\hat{\cal U}{\bf
P}_n \ ,
\label{T_approx}
\end{equation}
with the operator ${\hat G}^{(V)}$ defined by matrix elements
$\left<x,k\right|{\hat G}^{(V)}\left|x',m\right>=
\delta_{km}G_m^{(V)}(x,x')$.
Applying then Eqs.~(\ref{G_mn-sol}), (\ref{hatK}), and (\ref{T_approx}) to
Eq.~(\ref{Cond-mode}) we arrive at the following expression for the
impurity-averaged conductance,

\begin{eqnarray}
\left<g(L)\right>=-&&\frac{4}{L^2}\!\sum_{n=1}^{\infty}
\!\int\!\!\!\!\int_L\!\!dx\,dx'\!\left[
\Big<\frac{\partial G_{nn}(x,x')}{\partial x}
\frac{\partial G_{nn}^*(x,x')}{\partial x'}\Big> \right.
\nonumber \\
+&&
\frac{{\cal Q}}{d}\sum_{m=1\atop(m\neq n)}^{\infty}
\int_Ldy\, \left.
\Big<{G_m^{(V)}}^*(x,y)\frac{\partial}{\partial x}G_m^{(V)}(x,y)\Big>
\Big<G_{nn}(y,x')\frac{\partial}{\partial x'}
{G_{nn}}^*(y,x')\Big>\right] \ .
\label{G(L)_AV}
\end{eqnarray}

At this point it is useful to dicuss the spectral properties of the
quantum-mechanical system governed by the equation (\ref{1Deq}). First, the
term $\langle {\hat{\cal T}}_n\rangle=\Delta k_n^2-i/\tau_n^{(\varphi)}$
which modifies the initial spectrum $\{k_n^2\}$ can be readily calculated.
In the limit (\ref{weakscat}) an explicit form of the function
$W(\bbox{r})$ is not so important, and we obtain from Eq.~(\ref{T_approx})

\begin{mathletters} \label{renorm_spec}
\begin{equation}
\Delta k_n^2=\frac{{\cal Q}}{D}\sum_{m=1\atop (m\neq n)}^{\infty}
{\cal P}\!\!\!\int\limits_{-\infty}^{\infty}\frac{dq}{2\pi}
\frac{\widetilde{W}(q+k_n)}{k_m^2-q^2} \ ,
\label{corr_k2}
\end{equation}
\\[-2\baselineskip]
\begin{equation}
\frac{1}{\tau_n^{(\varphi)}}=\frac{{\cal Q}}{D}\sum_{m=1 \atop
(m\neq n)}^{N_c}\frac{1}{4k_m}\left[\widetilde{W}(k_n-k_m)
+\widetilde{W}(k_n+k_m) \right] \ .
\label{atten_n}
\end{equation}
\end{mathletters}
In Eq.~(\ref{corr_k2}) the symbol $\cal P$ stands for principal value,
$\widetilde{W}(q)$ is the Fourier transform of the correlation
function $W(\bbox{r}-\bbox{r}')$ over $y$, $y'$ and $x-x'$. The summation
in Eq.~(\ref{atten_n}) is restricted by the number $N_c=[k_FD/\pi]$ of
{\em conducting channels} (extended waveguide modes), because only for
$n\leq N_c$ the disorder-averaged diagonal in $n$ Green matrix $\big<{\hat
G}^{(V)}\big>$ in Eq.~(\ref{T_approx}) is essentially complex. The real
addition (\ref{corr_k2}) to the primordial mode energy $k_n^2$ is
small under conditions Eq.~(\ref{weakscat}), so that it can be omitted. At
the same time, the ``level broadening'' $1/\tau_n^{(\varphi)}$, which can
be interpreted as the inverse phase-breaking time for the $n$-th mode
state, relates to the mean level spacing as $N_c/k_F\ell$ and thus cannot
be omitted in the framework of the weak scattering approximation in
general. Just the addition (\ref{atten_n}) to $k_n^2$ is of crucial
importance for the further analysis.

We emphasize that the level broadening (\ref{atten_n}) implies the presence
of other extended modes with $m\neq n$ in the conductor. For extremely
narrow strips with $N_c=1$ the imaginary term is not present in the
renormalized mode spectrum as the sum (\ref{atten_n}) contains no terms
in this case. Then the system should exhibit true one-dimensional
properties.  Specifically, the electrons can be transferred within two
regimes only, {\em ballistic} and {\em localized}, and the conductance of
such a wire is decreased {\em exponentially} with the length $L$ exceeding
the {\em localization length} $\xi_1=4L_b\sim\ell$, the quadruple Born
backscattering length.\cite{MakTar98}

With increasing the conductor width, as soon as the wire ceases to
be single-mode ($N_c\geq 2$), the situation changes drastically. The
$n$-th mode spectrum is modified jointly by both the potentials $V_n(x)$
and $\Delta{\hat{\cal T}}_n$, and acquires the level broadening 
(\ref{atten_n}). We thus come to the necessity of analyzing the 
condition of (one-dimensional!) localization in {\em lossy} media, though 
{\em no inelastic scattering} was initially involved in the problem. The 
appropriate comprehensive theory is beyond the scope of the short article
and will be given in more extensive publication.\cite{FreilYurk94}

Here we emphasize that in studying spectral properties of a system governed
by Eq.~(\ref{1Deq}) one should clearly distinguish between the {\em direct}
intramode scattering due to the {\em local} potential $V_n(x)$ and {\em
indirect} intramode scattering due to the {\em operator} potential
${\hat{\cal T}}_n$. The intramode potential $V_n(x)$ gives rise to the
coherent localization effect, just as in the case with $N_c=1$,
Ref.~\onlinecite{MakTar98}. This potential causes local (in $x$) {\it
elastic} $n\to n$ transitions, so that the effect is purely interferential.
Meanwhile, from Eqs.~(\ref{Tn}), (\ref{T_approx}) it follows that the
potential ${\hat{\cal T}}_n$ can also be associated with the $n\to n$
scattering, but via all the other modes, i.e. except for the $n$-th mode.
Pictorially this can be thought of as diffusion in the mode space with
returning to the initial mode.

It is justified thus to regard the operator potential ${\hat{\cal T}}_n$
just as governing the {\em intermode} scattering within the effectively
``single-mode'' problem (\ref{1Deq}). This scattering can lead both to the
coherent localization (due to the potential $\Delta{\hat{\cal T}}_n$) and
to the uncertainty of a mode state due to the term
$i/\tau_n^{(\varphi)}$ arising from strong complexity of the trial
functions $G_m^{(V)}(x,x')$ at $m\leq N_c$. This duality of the
intermode scattering, especially the appearance of the ``phase breaking''
term $i/\tau_n^{(\varphi)}$ in spite of the scattering due to the
potential ${\hat{\cal T}}_n$ being effectively {\em intramode} (i.e., at
first sight elastic), has a clear physical explanation. It certainly
results from {\em probabilistic} nature of
electron transitions through intermediate mode states $m\neq n$
(intrinsic to the potential ${\hat{\cal T}}_n$) with the mode energies {\em
different} from $k_n^2$. This {\em hidden inelasticity} is exactly the
reason for strong complexification of the quasi-particle spectrum
(\ref{renorm_spec}).

In the final stage we discuss the role of the intermode scattering in the
whole range of the conductor length by estimating the Born scattering rate 
$1/\tau_n^{(\cal T)}$ which determines the fundamental frequency of the 
states presumably localized by the 1D random potential $\Delta{\hat{\cal 
T}}_n$. Estimation of the operator norm $\|\Delta{\hat{\cal T}}_n\|^2$ with 
the use of Eq.~(\ref{T_approx}) yields

\begin{equation}
\frac{\tau_n^{(\varphi)}}{\tau_n^{(\cal T)}}\sim
\frac{1}{\cos^2\vartheta_n}\text{min}\left(1,\frac{L/D}{k_F\ell}\right) \ ,
\label{estim_loc}
\end{equation}
where $\vartheta_n$ is a ``sliding angle'' of the mode $n$ with respect to
the $x$-axis, $|\sin\vartheta_n|=n\pi/k_FD$. The level broadening for an
$n$-th mode exceeds the level spacing provided the wire
is not extremely stretched along the $x$-axis, i.e. if the length $L$ does
not fall into the interval

\begin{equation}
L\gg Dk_F\ell\sim N_c\ell \ .
\label{quasi1D}
\end{equation}
Yet even within this interval the level spacing $1/\tau_n^{(\cal T)}$ due
to the potential $\Delta{\hat{\cal T}}_n$ {\em cannot exceed} the level
broadening $1/\tau_n^{(\varphi)}$. Consequently it is useless to seek the
traditional interferential localization at any length of the multi-mode
($N_c\geq 2$) conductor.

To illustrate the above statement we find the average conductance
(\ref{G(L)_AV}) for different lengths in the relatively easy case $N_c\gg
1$. The exact mode function $G_{nn}$ can be obtained from the equation

\begin{equation}
G_{nn}(x,x')=G_{nn}^{(0)}(x,x')+\Big(\hat G_{nn}^{(0)}
\Delta{\hat{\cal T}}_n\hat G_{nn}\Big)(x,x') \ ,
\label{Gnn_Dyson}
\end{equation}
which stems directly from Eq.~(\ref{1Deq}), where the ``unperturbed''
function $G_{nn}^{(0)}(x,x')$ obeys the equation (\ref{1Deq}) with
$\Delta{\hat{\cal T}}_n=0$. With the estimate (\ref{estim_loc}) taken into
account one can solve Eq.~(\ref{Gnn_Dyson}) perturbatively in
$\Delta{\hat{\cal T}}_n$. In doing so the addition to the conductance
emerges that is similar to the second term $\left<g^{(2)}(L)\right>$ on the
r.h.s. of Eq.~(\ref{G(L)_AV}), but proportional to higher degree of the
small interaction strength ${\cal Q}$.  The potential $\Delta{\hat{\cal 
T}}_n$ can thus be removed from Eq.~(\ref{1Deq}) and the intermode 
scattering taken into account through the dephasing rate 
$1/\tau_n^{(\varphi)}$ and the term $\left<g^{(2)}(L)\right>$. The 
potential $V_n(x)$, though different from $\Delta{\hat{\cal T}}_n$ by its 
physical meaning, can as well be removed from Eq.~(\ref{1Deq}) because of 
the relative smallness of its norm, $\langle\|\hat V_n\|^2\rangle 
/\langle\|\Delta{\hat{\cal T}}_n\|^2\rangle\sim N_c^{-1}$. Then the Green 
function $G_{nn}$ can be replaced in Eq.~(\ref{G(L)_AV}), to the main 
approximation in $N_c^{-1}\ll 1$, by its ``unperturbed'' expression

\begin{equation}
G_{nn}^{(0)}(x,x')=\frac{1}{2ik_n}
\exp\Big\{\big[ik_n-1/(\ell\cos\vartheta_n)\big]|x-x'|\Big\} \ ,
\label{Gnn^0}
\end{equation}
which nonetheless includes the most of inter-mode-scattering effects.

As to the functions $G_m^{(V)}(x,y)$ in Eq.~(\ref{G(L)_AV}), at $L\ll
N_c\ell$ we can put the potential $V_m(x)\equiv 0$ since the $m$-th
mode localization length found from Eq.~(\ref{GVn}) with the use of the
method of Ref.~\onlinecite{MakTar98} is
$
\xi_m=\frac{16\pi}{3}N_c\ell\cos^2{\vartheta_m}/\widetilde{W}_x(2k_m)
\sim N_c\ell
$.
In this case the second term in Eq.~(\ref{G(L)_AV}) turns out to be $-1/8$
of the first one, i.e. not parametrically small. Yet in the limit
(\ref{quasi1D}) all the functions $G_m^{(V)}(x,y)$ are localized, and
therefore the second term in Eq.~(\ref{G(L)_AV}) is negligibly small.

Basing on the above arguments we arrive at the following asymptotic
expressions for the conductance (\ref{G(L)_AV}), disregarding
weak-localization corrections governed by the intra-mode potentials
$V_n(x)$,

\begin{eqnarray}
 {\rm i)}\   L<\ell\ :&&\qquad
\left<g(L)\right>\approx N_c \ ;\nonumber\\[3pt]
 {\rm ii)}\   \ell\ll L\ll N_c\ell \ :&&\qquad
\left<g(L)\right>\approx \case{7\pi}{32}N_c\ell/L\gg 1 \ ;
\label{diffusive}\\[3pt]
 {\rm iii)}\  N_c\ell\ll L \ :&&\qquad
\left<g(L)\right>\approx \case{\pi}{4}N_c\ell/L\ll 1 \ .\nonumber
\end{eqnarray}
The result given in Eq.~(\ref{diffusive}) allows to distinguish three
regimes of charge transport in multi-mode conductors, none of them
localized in the anticipated sense. Regime (i) corresponds to entirely
ballistic transport, both from semiclassical and quantum standpoints. In
regimes (ii) and (iii) the semiclassical motion should be regarded as
diffusive. The difference between them is that in regime (ii) all the 
mode states could be considered extended in the absence of the intermode 
scattering, whereas in regime (iii) they all would be localized due to the
potentials $V_n(x)$. In both diffusive regimes (ii) and (iii) the
conductance exhibits purely ohmic (inversely proportional to $L$) 
behaviour, but with different (classic) diffusion coefficients. Note that 
just in regime (iii), when all the {\em trial states} would be localized if 
the inter-mode scattering was disregarded, the result is exactly reproduced 
given by the classical kinetic theory. No exponential decay of the
conductance appears at any length and width of the system provided
$N_c\geq 2$.

To conclude, the $T=0$ conductance of a 2D finite-size disordered metal
strip was calculated. The interferential localization was shown to manifest
itself strongly only for single-mode, i.e. purely 1D, conductors. In 
commonly examined square-shaped multi-mode samples the electron transport 
is diffusive as long as $L\gg\ell$, the semiclassical mean free path. For 
any extended (propagating) mode in a multi-mode strip all the other 
extended modes can be thought of as an effective phase-breaking reservoir 
destroying quantum interference and hence the exponential localization.

{\bf Acknowledgments --}\ \
The author is very grateful to N.~M.~Makarov for stimulating discussions
and to A.~V.~Moroz for help in the interpretation and presentation of the
results.

%%%%%%%%%%%%%%%%%%%%%%%%%%%%%%%%%%%%%

\end{document}